\documentclass[aps,pre,twocolumn,superscriptaddress,showpacs,amsmath,amssymb,floatfix]{revtex4}
\usepackage{float}
\usepackage{epsfig}
\usepackage{amsmath}
\usepackage{amssymb}
\usepackage{bm}
\usepackage{color}
\usepackage{graphicx}
\usepackage{graphicx,rotating}

\begin{document}

\author{Benjamin J. Galow}
\affiliation{Max-Planck-Institut f\"{u}r Kernphysik, Saupfercheckweg 1,
69029 Heidelberg, Germany}
\author{Jian-Xing Li\footnote{On leave from the College of Physics Science and Information Engineering, Hebei Normal University, Shijiazhuang 050016, China.}}
\affiliation{Max-Planck-Institut f\"{u}r Kernphysik, Saupfercheckweg 1,
69029 Heidelberg, Germany}
\author{Yousef I. Salamin}
\affiliation{Max-Planck-Institut f\"{u}r Kernphysik, Saupfercheckweg 1,
69029 Heidelberg, Germany}
\affiliation{Department of Physics, American University of Sharjah, POB 26666, Sharjah,
United Arab Emirates}
\author{Zolt\'an Harman}
\affiliation{Max-Planck-Institut f\"{u}r Kernphysik, Saupfercheckweg 1,
69029 Heidelberg, Germany}
\affiliation{ExtreMe Matter Institute EMMI, Planckstrasse 1,
64291 Darmstadt, Germany}
\author{Christoph H. Keitel}
\affiliation{Max-Planck-Institut f\"{u}r Kernphysik, Saupfercheckweg 1,
69029 Heidelberg, Germany}

\pacs{52.38.Kd, 37.10.Vz, 42.65.-k, 52.75.Di, 52.59.Bi, 52.59.Fn, 41.75.Jv,
87.56.bd}

\title{High-quality multi-GeV electron bunches via cyclotron autoresonance}

\begin{abstract}

Autoresonance laser acceleration of electrons is theoretically investigated using circularly polarized focused Gaussian pulses.
Many-particle simulations demonstrate feasibility of creating over 10-GeV electron bunches of ultra-high quality
(relative energy spread of order $10^{-4}$), suitable for fundamental high-energy particle physics research.
The laser peak intensities and axial magnetic field strengths required are up to about $10^{18}$~W/cm$^2$ (peak power $\sim10$~PW) and 60~T, respectively.
Gains exceeding 100~GeV are shown to be possible when weakly focused pulses from a 200-PW laser facility are used.

\end{abstract}

\maketitle

\section{Introduction}

Particle accelerators are an indispensable tool to explore the fundamental laws of nature and are widely used for medical and industrial applications.
At the frontier of accelerator technology is the Large Hadron Collider (LHC), a gigantic circular machine of 27 km total circumference \cite{lhc}.
The need to control the size and cost of building such machines have kept alive the quest for alternative means to accelerate particles.
Over the past decade, laser plasma-based acceleration has emerged as a promising candidate \cite{snaveley2000,mackinnon2001,karsch2003,romagni2005,hegelich2006,schwoerer2006,robson2007}.
In particular, laser wakefield acceleration of electrons \cite{mangles2004,leemans2004,faure2004} has undergone rapid
development. Stable and reproducible beams have been realized \cite{faure2006} and particle kinetic energies
at the GeV level have been reached \cite{leemans2006}. Furthermore, the creation of a plasma wave from interaction with
a highly energetic electron beam as a driver allows for doubling the kinetic energy of the accelerated particles within
a meter-scale plasma wakefield accelerator \cite{blumenfeld2007,leemans2009,leemans2009phystoday}.

The advent of quasi-static magnetic fields \cite{lanl1, lanl,dresden,florida1,florida2} of durations up to seconds, with strengths as high
as 100 Tesla, suggests vacuum autoresonance laser acceleration (ALA) (see \cite{loeb1986,sal2000} and references therein)
as a further potential alternative to conventional acceleration. The ALA mechanism employs a static magnetic field oriented
along the propagation direction of the laser. Thus, the underlying concept of ALA stems from the realization that an electron
continues to absorb energy from a circularly polarized laser field if it is launched in cyclotron autoresonance with it. For the case of laser
fields described by plane-waves \cite{sal2000} resonance is essentially between the Doppler-shifted laser frequency seen
by the electron and the cyclotron frequency of the electron around the lines of the applied static magnetic field.
Feasibility of post-acceleration of electrons to kinetic energies of about three times their initial energies has also been
theoretically investigated, employing continuous-wave CO$_2$ laser fields described within the paraxial approximation \cite{hirshfield2000}.

\begin{figure}[b]
\includegraphics[width=8cm]{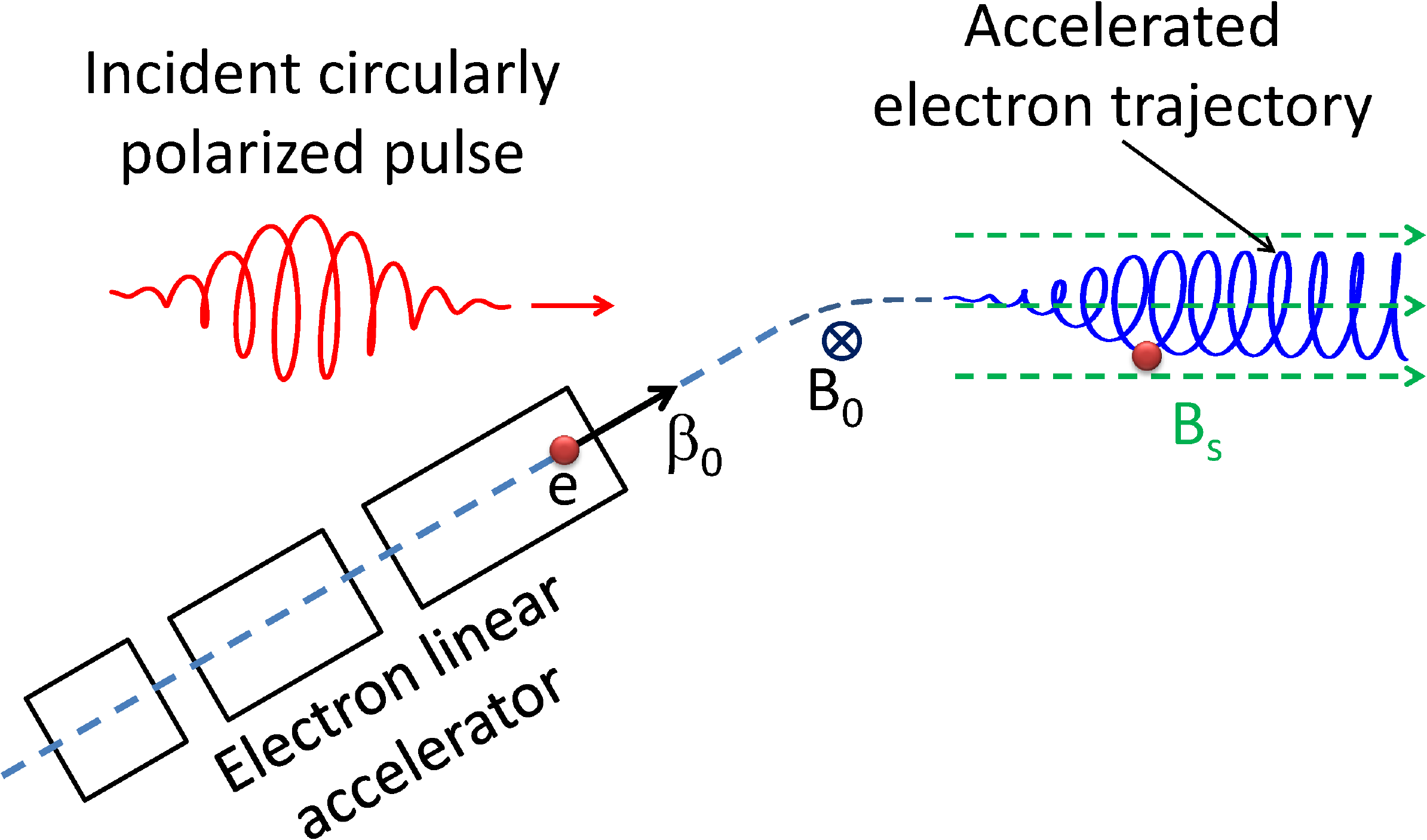}
\caption{(Color online) A schematic showing the vacuum ALA configuration. A linear accelerator (LINAC) pre-accelerates the electron, labeled by $e$ and the velocity
vector $\bm{\beta}_0$, to 50~MeV and a magnetic field $B_0$ bends its trajectory slightly for injection axial to the combination of incident laser
pulse and axial uniform magnetic field of strength $B_s$.}\label{schematic}
\end{figure}

In this paper, the ALA configuration  (see Fig. \ref{schematic}) is investigated over a wide range of laser and magnetic field parameters.
Our results, stemming from single- and many-particle calculations which employ pulsed and focused laser fields, indicate electron energy gains
of several hundred (even thousand) times the initial injection energy. For magnetic field strengths below 60~Tesla, energy gains
in excess of 10~GeV are shown to be possible. It is shown that the gains are attained over distances less than 10~m, and from a near-infrared laser system
of peak intensity $<10^{18}$~W/cm$^2$ (peak power $\sim10$~PW). Many-particle simulations also demonstrate that an electron bunch of high quality (relative energy spread on the $10^{-4}$ level)
may be obtained, taking Coulomb particle-particle repulsions into account
in simulations at densities of $10^{15}$~cm${}^{-3}$.  Gains in excess of 100~GeV are also shown to be possible anticipating
a 200-PW laser system, like it might be realized by the Extreme Light Infrastructure (ELI) \cite{eli}.
Our work is motivated by currently feasible magnetic fields of strength of the order of 100 Tesla \cite{lanl1, lanl} and anticipates continued progress in high magnetic field research.

\section{Basic equations}

Classical electron motion in the presence of electromagnetic fields $\bm{E}$ and $\bm{B}$ is governed by the
Lorentz-Newton equations of motion, namely,

\begin{equation}
    \frac{d\bm{p}}{dt}=-e (\bm{E}+c\bm{\beta}\times\bm{B})\quad \text{and}\quad \frac{d\mathcal{E}}{dt}=-ec\bm{\beta}\cdot\bm{E},
\end{equation}
which describe time evolution of the particle's relativistic momentum $\bm{p}=\gamma mc\bm{\beta}$ and energy $\mathcal{E}=\gamma mc^2$, respectively. In the above, $e$ is the magnitude of the charge of the electron and $m$ is its rest mass, $c$ is the speed of light in vacuum, $\bm{\beta}$ is the particle's
velocity scaled by $c$, $\gamma=(1-\beta^2)^{-1/2}$ and SI units are used throughout. Results to be presented below are based on solving these equations numerically
for single- and many-particle systems. In the solutions, the electron is assumed to be overcome by the front of the pulse at $t=0$ at the
origin of a coordinate system whose $z-$axis is oriented along the direction of pulse propagation.

In earlier calculations of autoresonance acceleration the laser fields were modeled as plane-waves of infinite extension in space and time \cite{sal2000},
or as those of a continuous beam within the paraxial approximation \cite{hirshfield2000}. The plane-wave-based calculations have led to the realization
of the resonance condition, to be recalled below, and have shown that an electron stands to gain more energy from circularly polarized light than from
light of the linear polarization variety. Recall that the polarization vector of a circularly polarized plane wave rotates about the direction of
 propagation at the angular frequency of the wave, while its field strength remains constant. Thus, if the initial conditions are such that the electron
cyclotron frequency matches the Doppler-shifted frequency (sensed by the electron) of the circularly polarized fields, the electron will subsequently {\it surf}
on the wave and continue to absorb energy from it. Hence, calculations in this paper will employ circularly polarized laser fields, obtained by the superposition
 of two linearly polarized fields, with perpendicular polarization vectors and a $\pi/2$-phase difference \cite{sal2007}. A pulse shape is introduced by multiplying
the fields by the Gaussian envelope $\exp(-\eta^2/2 \sigma^2)$, where $\eta=\omega t-k z$ is the phase variable, $k=2\pi/\lambda$ is the wavenumber,
$\sigma=\omega \tau/(2\sqrt{2\ln2})$ is the envelope's full-width-at-half-maximum, and $\tau$ is the pulse duration (temporal full-width-at-half-maximum).
 For the field amplitudes, a generalized Lax series representation (in powers of the diffraction angle $\epsilon=\lambda/\pi w_0$, where $w_0$ is the beam's waist
radius at focus) will be adopted \cite{lax,davies,sal2007}. Thus, the fields of the ALA scheme may be written as (see \cite{sal2007} for definitions of the symbols and more details)

\begin{eqnarray}
\label{E} \bm{E} &=& e^{-\eta^2/2\sigma^2}\left\{\left[E_x\hat{\bm{e}}_x+E'_x\hat{\bm{e}}_y\right]+\left[E_y\hat{\bm{e}}_y+E'_y\hat{\bm{e}}_x\right]\right.\nonumber\\
		  & &	\left.+\left[E_z+E'_z\right]\hat{\bm{e}}_z\right\},
\end{eqnarray}
and
\begin{eqnarray}
\label{B} \bm{B} &=& e^{-\eta^2/2\sigma^2}\left\{\left[B_y\hat{\bm{e}}_y-B'_y\hat{\bm{e}}_x\right]\right.\nonumber\\
		 & & \left.+\left[B_z-B'_z\right]\hat{\bm{e}}_z\right\}+B_s\hat{\bm{e}}_z,
\end{eqnarray}
where the primed components follow from the unprimed ones by letting $x\leftrightarrow y$ and adding a phase-shift of $-\pi/2$.

\section{The fields}

\begin{figure}[b]
\includegraphics[width=8cm]{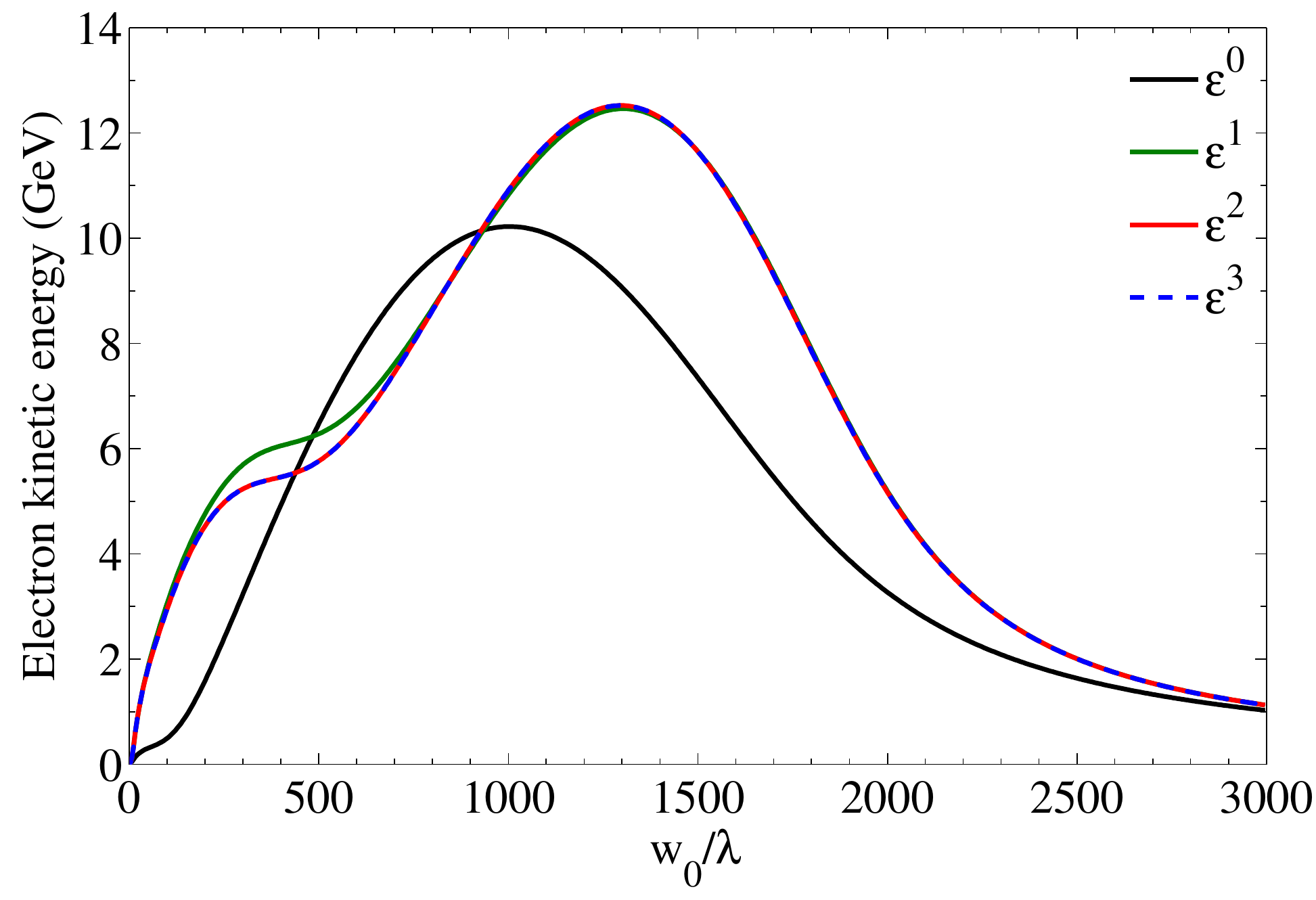}
\caption{(Color online) Electron exit energy gain as a function of the laser beam radius at focus, employing field representations of different orders in the diffraction angle. The electron is injected along the $z$-axis with 50 MeV initial kinetic energy and initial location at the origin of coordinates $x=y=z=0$. The laser system parameters are: $\lambda=1~\mu$m, power $P=10$ PW, and pulse duration $\tau=25$ fs.
The trajectory and the excursion distance of the electron
for the set of parameters leading to maximal kinetic energy gain are shown in Fig. \ref{traj}(a).
}
\label{order}
\end{figure}

To decide the order of the correction terms, beyond the paraxial approximation, which ought to be retained in the various field expressions, simulations have been performed
for a single electron injected axially with 50 MeV initial kinetic energy. The electron's exit energy gain as a function of
the pulse waist radius at focus has been analyzed, when terms up to ${\cal O}(\epsilon^n)$, where $n=0, 1, 2, 3$, are employed in modeling the laser fields (see Fig. \ref{order}).
The simulation results for terms of highest order $\epsilon^2$ and $\epsilon^3$ coincide, demonstrating that terms of order higher than
$\epsilon^2$ may be dropped, as expected, since $\epsilon\ll1$. 

Due to the large focus radius, the longitudinal component
influences the ALA dynamics only negligibly and the energy gain results essentially from interaction with the transverse field component. This
is shown in analytical calculations in \cite{sal2000} for purely transverse fields.
For the parameters used (see caption of Fig.~\ref{contour}), the exit
energy gain peaks for a waist radius at focus $w_0\sim1295\lambda$.
The maximum electron exit energy gain attained in this case is $K\sim12.5$~GeV.
The peak is reached for a focus large enough to allow for optimal autoresonance to occur, but still tight enough to guarantee a sufficiently
strong field ($E_0\propto1/w_0$).

\section{Cyclotron Autoresonance}

\begin{figure}[t]
\includegraphics[width=8cm]{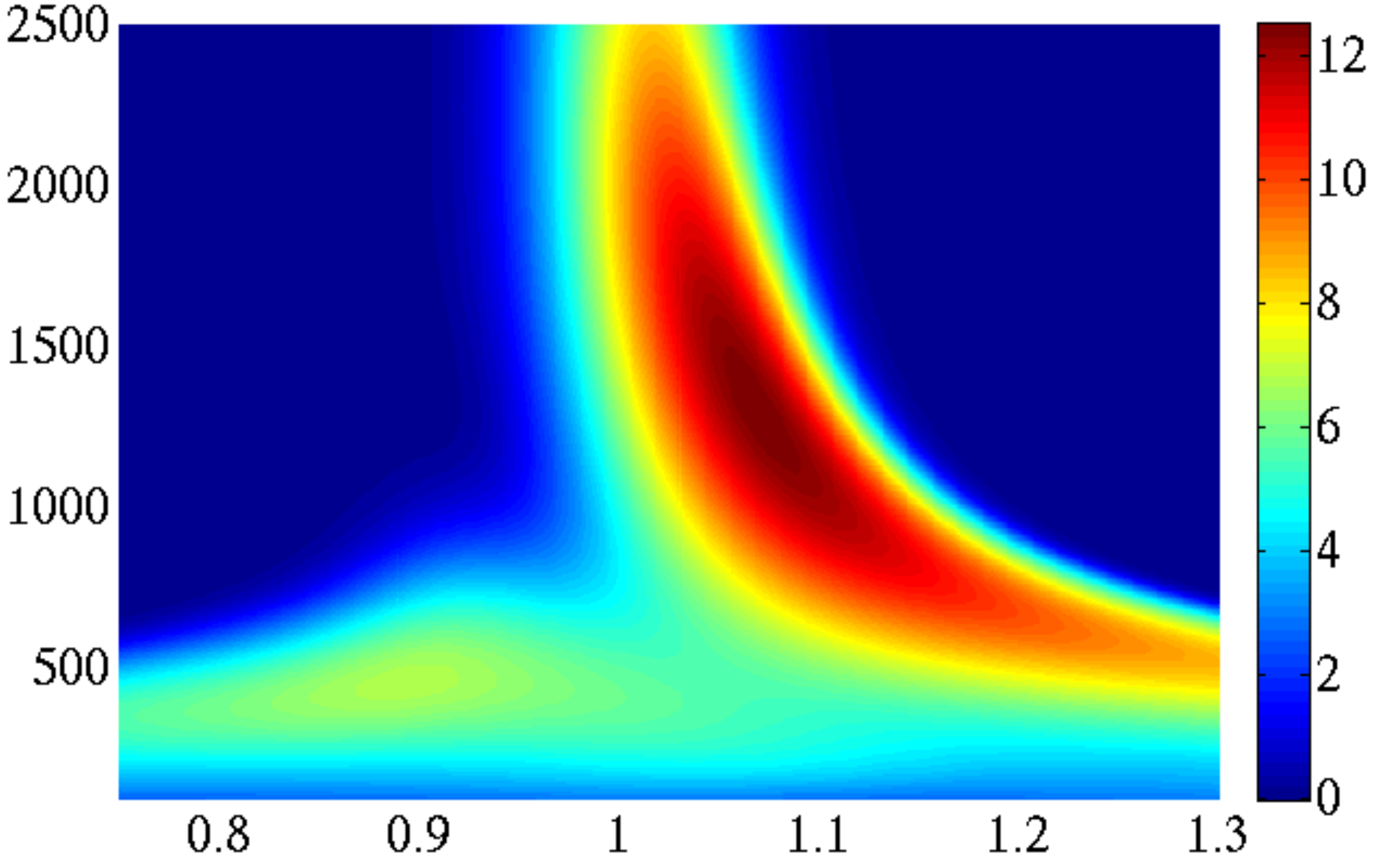}
\begin{picture}(0,0)(100,10)
\put(-25,0){$B_s/B_{s0}$}
\put(100,39){\begin{sideways}Exit energy gain (GeV)\end{sideways}}
\put(-140,75){\begin{sideways}$w_0/\lambda$\end{sideways}}
\end{picture}
\caption{(Color online) Contour plot of the exit energy gain as a function of the beam radius at focus $w_0$ and the external
static magnetic field $B_s$.
The laser and electron injection parameters are the same as in Fig. \ref{order}. }
\label{contour}
\end{figure}
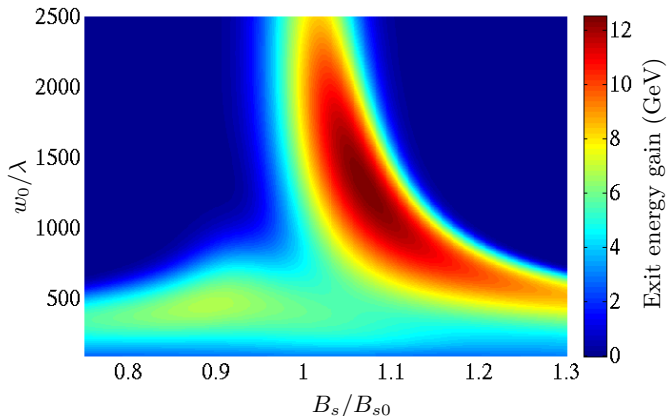

From the plane-wave calculations \cite{sal2000} we learn that, for autoresonance to occur, the uniform magnetic field, to be applied axially and added to the laser magnetic field in the equations of motion, should be calculated from
\begin{equation}
    B_{s0}=\frac{m \omega}{e}\sqrt{\frac{1-\beta_0}{1+\beta_0}},
\label{condition}
\end{equation}
where $\omega$ is the laser frequency, and $\beta_0$ is the initial speed of the electron scaled by $c$. Introduction of
a pulse-shape and a tight focus is expected to modify the resonance condition (\ref{condition}) and render
it approximate at best, thus leading to a slight deterioration
in the electron-beam quality. To investigate this issue, the applied axial magnetic
field $B_{s0}$ is replaced by $B_s$ and the parameter space, spanned by $w_0/\lambda$ and $B_s/B_{s0}$, is scanned for optimum exit energy gains. The
results are displayed in Fig.~\ref{contour}; a contour plot of the exit energy gain vs. both the beam waist radius at focus $w_0$ and the employed static
magnetic field $B_s$. The plot shows clearly a region in parameter space for which the exit
energy gain is optimal. The exit energy gain is not sensitive to small fluctuations in the ALA configuration parameters.
For example, an energy gain of about 11~GeV may be realized for waist radii $w_0$ in the range extending roughly from 950$\lambda$ to 1750$\lambda$ and
a magnetic field strength $B_s$ in the approximate range 1.04 $B_{s0}$ to 1.13$B_{s0}$. The maximum energy gain of 12.5~GeV is reached
for $B_s=1.07B_{s0}\sim58$ T. Recent progress in quasi-static magnetic field research has achieved 45 T \cite{florida1,florida2,lanl}
over a distance of 0.225~m. Thus, the 58 T goal may not be too far-fetched, and the need to have such a magnetic field strength over 25
m may, in principle, be met by employing an assembly of such magnets.
However, such a stacking of magnets
will be experimentally difficult to realize and the generated fringe fields at the
interface between two magnets will probably have negative impact on the
energy resolution.

To continuously maintain a magnetic field strength of about 60 T over 10 m for one second would
require an average power consumption of 2.25 GW \cite{lanl}. However, a petawatt laser provides its energy 
in a pulsed way, such that we do not need to continuously maintain the magnetic field which will
significantly lower the average energy consumption as we will show subsequently. 
 For one relativistic
electron bunch at approx. the speed of light it takes about 33.3 ns to travel 10 m and, hence, for
ten bunches about 0.3 $\mu$s. This implies that at a laser repetition rate of 10 shots per second
the energy consumption of the ALA scheme corresponds to only 0.7 kJ based on pulsed
magnets. Since this value does not include the power consumption of the laser system and the
LINAC, it exceeds in total the power consumption of laser wakefield accelerators \cite{martins2010,welsh2012}.
Moreover, we want to emphasize that this estimate is based on the assumptions that the
magnets are able to operate at such short pulses and their power consumption scales linearly with
the magnets' pulse duration, which might be influenced by the presence of pedestals.

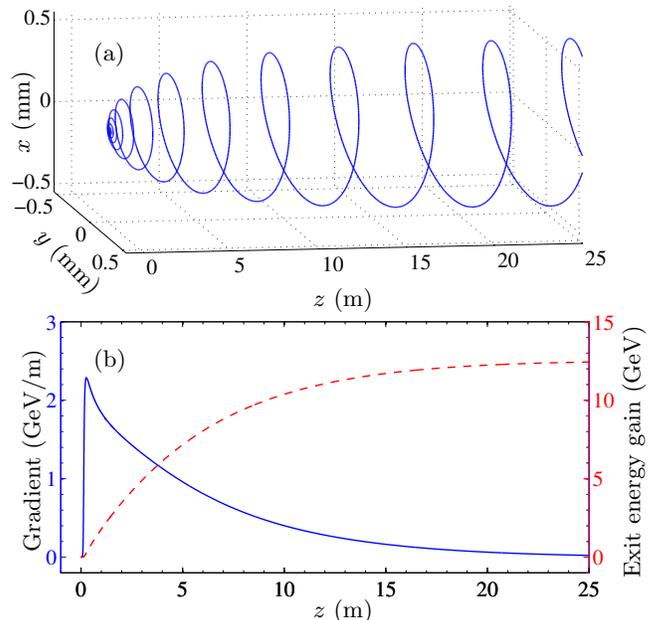
\begin{figure}[b]
\includegraphics[width=8cm]{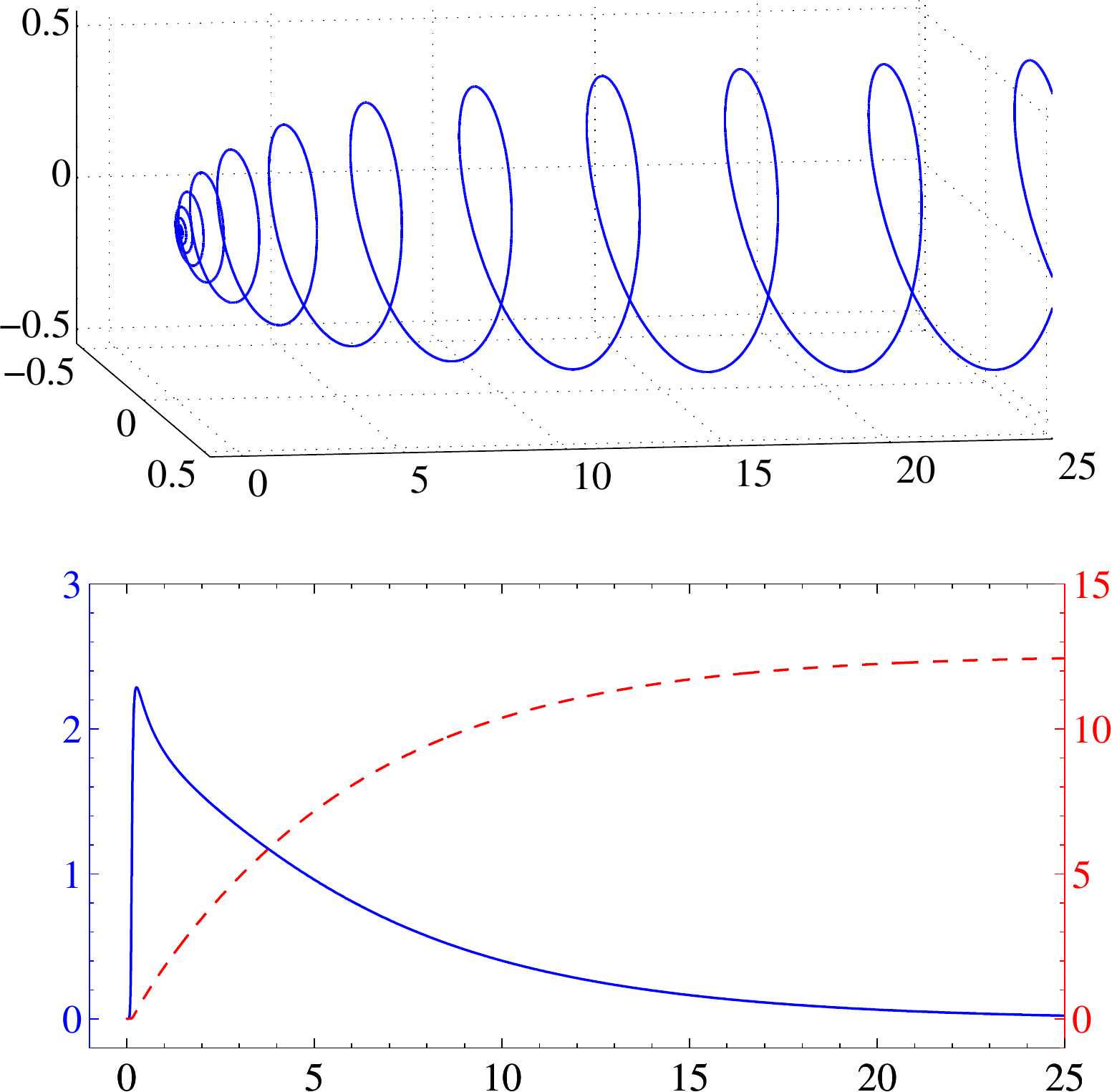}
\begin{picture}(0,0)(100,10)
\put(-17,120){$z$ (m)}
\put(-122,146){\begin{rotate}{-46} $y$ (mm) \end{rotate}}
\put(-131,178){\begin{sideways}$x$ (mm)\end{sideways}}
\put(-17,2){$z$ (m)}
\put(-100,213){(a)}
\put(-100,97){(b)}
\put(-127,27){\begin{sideways}Gradient (GeV/m)\end{sideways}}
\put(100,17){\begin{sideways}Exit energy gain (GeV)\end{sideways}}
\end{picture}
\caption{(Color online) (a) 3D trajectory of a single electron in autoresonance interaction with laser and static magnetic fields,
for $w_0\sim1295\lambda$ and $B_s=1.07B_{s0}$. (b) Acceleration gradient (blue line, left ordinate) and exit energy gain (red dashed line, right ordinate) vs. the axial excursion distance
for the same electron.
The laser peak intensity is $I_0\sim7.59\times10^{17}$ W/cm$^2$, and other parameters are the same as in Fig. \ref{contour}.}
\label{traj}
\end{figure}

\section{ALA dynamics}
\subsection{Single-particle calculations}

For further insight into the ALA dynamics, Fig. \ref{traj}(a) shows the 3D trajectory of a single electron, accelerated  using the optimal parameters of $w_0=1295\lambda$
and $B_s=1.07B_{s0}$. Note that the transverse dimensions are enlarged in Fig. \ref{traj}(a)
giving the impression that the trajectory is a helix with increasing radius. However, the radius of the helix hardly increases beyond the first 9 m of
axial excursion, over which the pitch of the helix increases rapidly, thus rendering the trajectory essentially linear.
In Fig. \ref{traj}(b) the
corresponding exit energy gain (red dashed line) is shown as a function of the axial excursion distance. After an initial (approximately linear) increase, the energy gain starts
to saturate with increasing excursion.
The accelerating phase seems to be
limited within the initial axial excursion, after which interaction with the transverse electric field components diminishes as the particle gets left behind the pulse.
For the set of parameters used (see caption of Fig. \ref{contour}) the 10-GeV level is reached after a 9.14-m axial excursion.
Moreover, the acceleration gradient $dG/dz=-e\bm{\beta}\cdot\bm{E}/\beta_z$ is plotted vs. the axial excursion $z$ (blue line).
The maximal acceleration gradient of 2.286~GeV/m is reached at $z = 0.2626$~m.
This exceeds the gradient of conventional LINACs (100~MeV/m) by more than one order of magnitude.
By contrast, wakefield accelerators reach gradients of 10-100~GeV/m~\cite{leemans2006}. However, it should be emphasized
that in single-stage wakefield accelerators the acceleration distance is typically $<1$~m, yielding energy spreads of
the order of $2\%$~\cite{leemans2004}.

\subsection{Many-particle simulations}

The single-particle calculations, whose results have been presented above, will now be supported by many-particle simulations.
Dynamics of a bunch of electrons injected along the $z$-axis into the ALA configuration, as well as
the beam properties of the accelerated electrons, are considered next. An ensemble of electrons,
considered to be {\it non-interacting} for now, randomly distributed within a volume of cylindrical
shape centered about the coordinate origin and oriented along the $z$-axis, is used to model an electron bunch,
 along the lines of our earlier work in \cite{sal-prl2,sal-pra2011,li}.
The incident laser pulse accelerates particles at the left end of the cylinder first, followed by particles farther
to the right.
The cylinder containing the electrons has a
 length $l_c=1$ mm and a radius $r_c=0.05$ mm. 
 
 The initial kinetic energy of the electrons follows a
normal distribution with mean value $K_0=50$~MeV and spread (standard deviation) $\Delta K_0=0.05$~MeV \cite{lee}. Such an electron bunch
 may be pre-accelerated using a short LINAC or a table-top betatron, and then guided by a magnetic
 field for axial injection (see Fig. \ref{schematic}).
Employing the laser system parameters of Fig. \ref{contour} and the optimal waist radius $w_0=1295\lambda$, the resulting exit energy gain distribution
 of an ensemble of 15000 electrons has a mean exit energy gain of $G_{{\rm exit}}=12.513$~GeV and a spread of
 $\Delta G_{\rm exit}=3.7$~MeV ($0.0293\%$).
 The transverse beam emittance amounts to $\approx 0.1$ $\pi$ mm mrad which compares well with what is obtained from
conventional accelerators \cite{lee}.

Dependence of the electron exit kinetic energy distribution on fluctuations in the initial kinetic
energy distribution has been studied. Employing a bunch with $K_0=50\pm0.5$~MeV
changes the mean exit energy gain to $G_{\rm exit}=12.296$~GeV, and its spread to $\Delta G_{\rm exit}=0.321$~GeV ($2.61\%$).

\subsection{Particle-particle interaction effects}

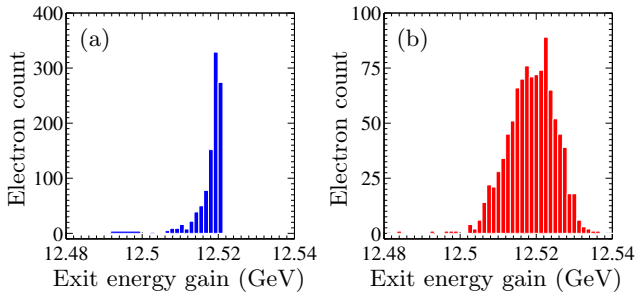
\begin{figure}[t!]
\includegraphics[width=8cm]{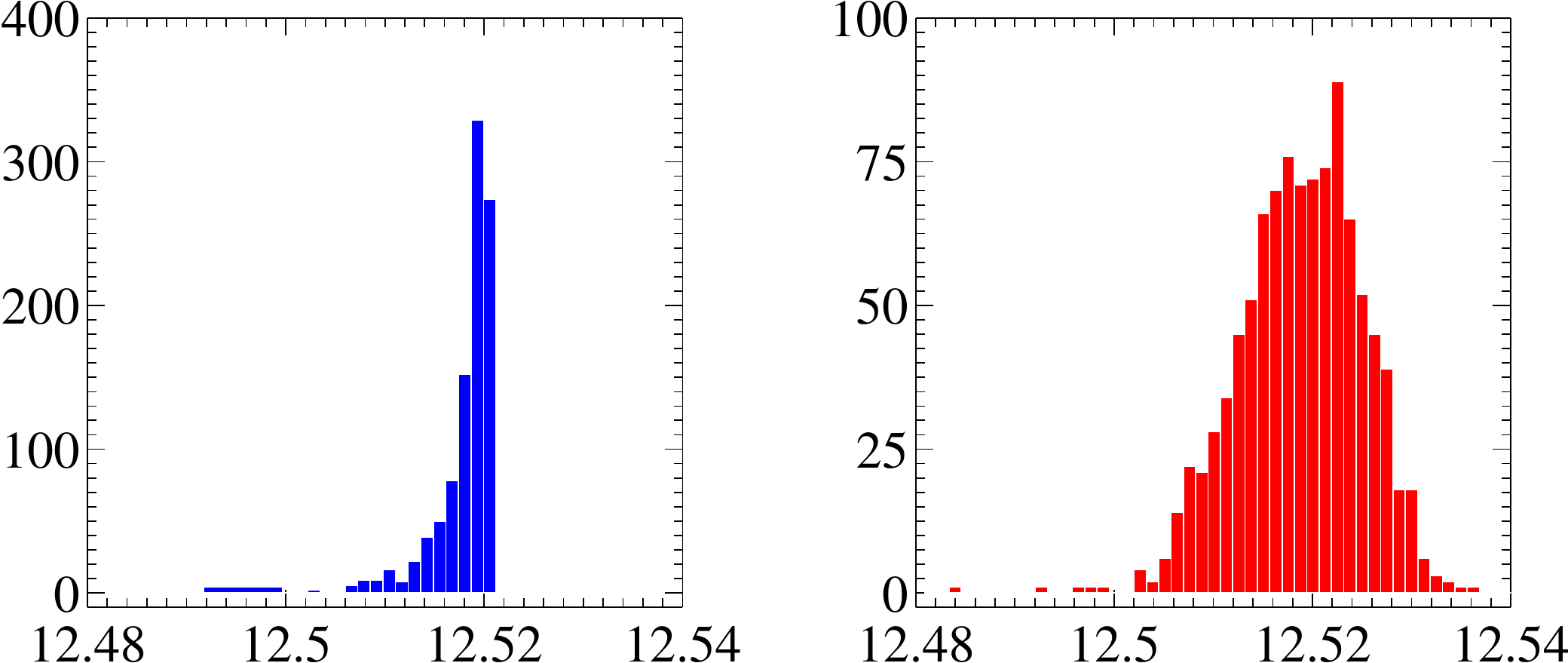}
\begin{picture}(0,0)(100,10)
\put(-113,91){(a)}
\put(7,91){(b)}
\put(-123,0){Exit energy gain (GeV)}
\put(-3,0){Exit energy gain (GeV)}
\put(-140,30){\begin{sideways}Electron count\end{sideways}}
\put(-18,30){\begin{sideways}Electron count\end{sideways}}
\end{picture}
\vskip2pt
\caption{(Color online) Distribution of the exit energy gain amongst 1000 (a) {\it non-interacting} and (b) {\it interacting} electrons in an ALA scheme.
 See Figs. \ref{contour} and \ref{traj} for the laser and injection parameters.}
\label{hist}
\end{figure}

To investigate the role of electron-electron interaction effects a suitable model has to be developed. A
conventional particle-in-cell scheme describes the interaction of a laser with an initially neutral plasma, and is not applicable over macroscopic
distances of several meters.
Therefore, further simulations have been performed employing a 1000-particle ensemble confined to a spatial volume similar to what has been used above,
but scaled to render the particle density the same as would be obtained from a $10^{10}$-particle bunch
(typical in conventional particle accelerators \cite{slac-pub}) with the Coulomb interactions turned on \cite{galow} and off.
The resulting exit energy gain distributions are shown in Fig.~\ref{hist}.
In the {\it non-interacting} ensemble case Fig.~4~(a) the mean exit energy gain amounts to $G_{\rm exit}=12.518$~GeV with a spread
of $\Delta G_{\rm exit}=3.17$~MeV ($0.0253\%$). For the {\it interacting} ensemble Fig.~4~(b) the energy spread approximately
doubles ($G_{\rm exit, 1000}^{\rm Coulomb}=12.519$~GeV, $\Delta G_{\rm exit, 1000}^{\rm Coulomb}=6.22$~MeV ($0.0497\%$)).
Furthermore, spatial spreading of the bunch (not shown here) is increased by a few percent as a result.

Since the relative velocities of electrons in the center-of-mass of the bunch are low ($\beta_{\rm rel.}\lessapprox10^{-3}$), higher-order relativistic particle-particle interaction
effects \cite{galow} can be neglected. To ensure that the reduced size of the ensemble does not play a role we performed
the simulations for an {\it interacting} ensemble of 500 particles at the same density. Bearing in mind the different random initial conditions,
a mean exit energy gain of $G_{\rm exit, 500}^{\rm Coulomb}=12.519$~GeV with a spread of $\Delta G_{\rm exit, 500}^{\rm Coulomb}=5.18$~MeV ($0.0420\%$) has been obtained, which is in good
agreement with the values given above for the 1000-particle ensemble. Hence, validity of the calculational method is confirmed and the
long-range interaction effects have been treated appropriately.

\section{Discussion}

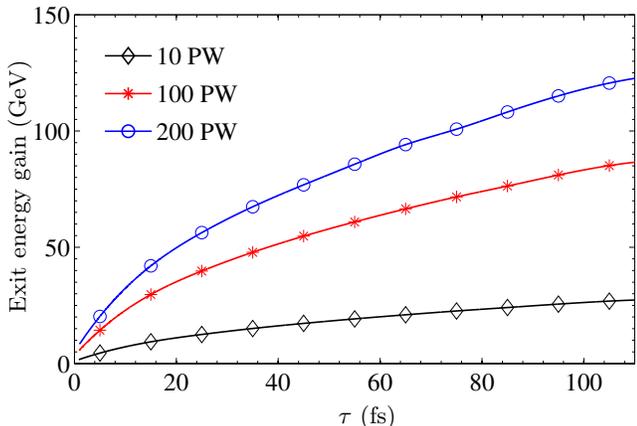
\begin{figure}
\includegraphics[width=8cm]{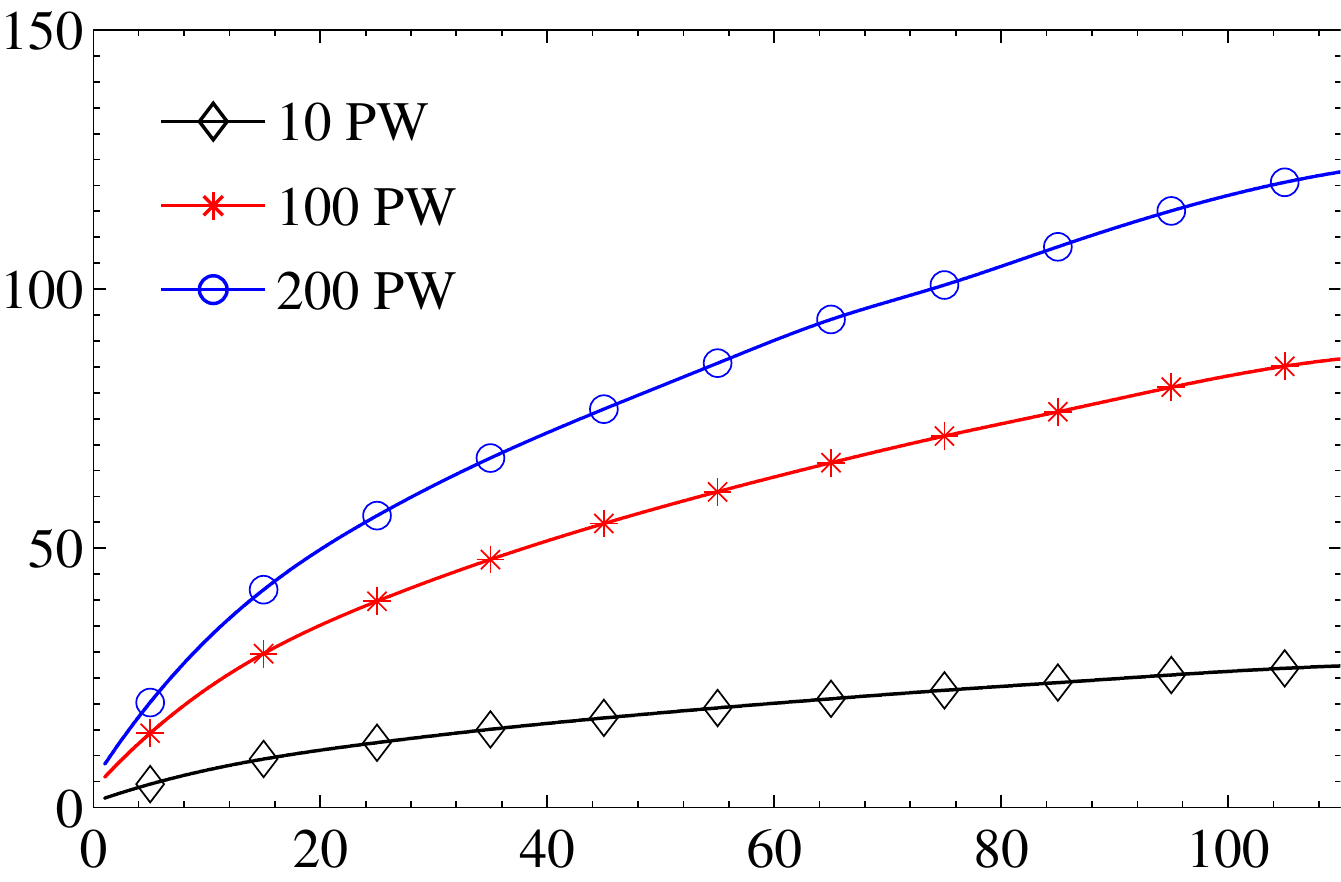}
\begin{picture}(0,0)(100,10)
\put(-15,0){$\tau$ (fs)}
\put(-140,42){\begin{sideways}Exit energy gain (GeV)\end{sideways}}
\end{picture}
\caption{(Color online) Exit electron energy gain vs. pulse duration for three laser powers.
Optimal magnetic field strengths and waist radii at focus (obtained along the lines of Fig. \ref{contour}, for every data point) have been used.
 The injection parameters are the same as in Fig. \ref{contour}.}
\label{future}
\end{figure}

The examples discussed thus far have been concerned essentially with a single set of laser parameters, soon to be available for laboratory experiments.
Even more powerful laser systems may be available in the near future \cite{eli}.
In search of parameter sets that may lead to much higher energy gains, useful for particle physics research, single-particle calculations
have been performed whose results are displayed in Fig. \ref{future}. The figure shows, e.g., that a gain of about 100~GeV may be reached employing a 75-fs pulse,
focused to $w_0=6800\lambda$ and derived from a 200-PW laser system (corresponding to a peak intensity $I_0\sim4.41\times10^{17}$ W/cm$^2$) \cite{eli}.
However, this will come at a price to be paid in terms of the size of such a facility, and a uniform magnetic field $B_s=55.3$~T to be maintained along around 400~m.
To lower the required magnetic field strength one may inject the electrons at a higher initial velocity, as can be seen from Eq.~(\ref{condition}).

For electrons accelerated to tens of GeV energy or above, the question of whether radiation reaction effects play a role may arise.
However, simulations based on the Landau-Lifshitz equation \cite{landau} have also been carried out which revealed that the electron dynamics is
only marginally influenced by radiation reaction. This agrees well with the finding that, at intensities below $10^{18}$~W/cm${}^2$,
radiation reaction effects are negligible~\cite{antonino}.

\section{Summary and conclusions}

It has been demonstrated, in single- and many-particle simulations, that electrons may be accelerated to multi-GeV
energies, if launched into cyclotron autoresonance with a circularly polarized laser pulse, and employing parameters for the laser and required uniform magnetic
field that are currently available, or under construction. Similar simulations have also been shown to lead to over 100-GeV electron energy gains from envisaged
laser pulses \cite{eli}. In all cases considered, the energy gradients exceed the known limits of conventional accelerators by at least one order of magnitude.
Dedicated many-particle simulations reveal ultra-low relative energy spreads $\Delta G/G$ of the order of $10^{-4}$ comparable with conventional accelerator and storage facilities
\cite{lee} and suitable for high-precision particle
physics experiments. 
However, we want to recall that the parameters used in this theoretical study
particularly for the employed magnetic field strengths over long distances are out of the scope of near-future
experiments.\\

\begin{acknowledgments}
BJG acknowledges discussions with T. V. Liseykina, A. Di Piazza and M. Tamburini, BJG and JXL acknowledge hospitality at the American University of Sharjah (UAE)
where part of this work was done, and YIS acknowledges support from the German Alexander von Humboldt Stiftung in Bonn.
The work of ZH has been supported by the Alliance Program of the Helmholtz Association (HA216/EMMI).
\end{acknowledgments}

\end{document}